\documentclass[twocolumn,preprintnumbers,amsmath,amssymb]{revtex4}
\input psfig.sty

\usepackage{graphicx}
\usepackage{dcolumn}
\usepackage{bm}

\newcommand{\be}{\begin{equation}}
\newcommand{\ee}{\end{equation}}

\begin{document}


\title{Alternative Dark Energy Models: An Overview}

\author{J. A. S. Lima$^{1,2}${\footnote{emails: limajas@astro.iag.usp.br, limajas@dfte.ufrn.br}}}

\address{$^{1}$Instituto de Astronomia, Geof\'{\i}sica e Ci\^encias
Atmosf\'ericas, 05508-900 S\~ao Paulo, SP, Brazil
\\
$^{2}$Departamento de F\'{\i}sica, Universidade Federal do Rio
Grande do Norte, 59072-970, Natal, RN, Brazil}


\begin{abstract}
\hspace{5.3cm}{\bf ABSTRACT}

\vspace{0.1cm} A large number of recent observational data
strongly suggest that we live in a flat, accelerating Universe
composed of $\sim$ 1/3 of matter (baryonic + dark) and $\sim$ 2/3
of an exotic component with large negative pressure, usually named
{\bf Dark Energy} or {\bf Quintessence}. The basic set of
experiments includes: observations from SNe Ia, CMB anisotropies,
large scale structure, X-ray data from galaxy clusters, age
estimates of globular clusters and old high redshift galaxies
(OHRG's). Such results seem to provide the remaining piece of
information connecting the inflationary flatness prediction
($\Omega_{\rm{T}} = 1$) with astronomical observations.
Theoretically, they have also stimulated the current interest for
more general models containing an extra component describing this
unknown dark energy, and simultaneously accounting for the present
accelerating stage of the Universe. An overlook in the literature
shows that at least five dark energy candidates have been proposed
in the context of general relativistic models. Since the
cosmological constant and rolling scalar field models have already
been extensively discussed, in this short review we focus our
attention to the three remaining candidates, namely: a decaying
vacuum energy density (or  ${\bf \Lambda(t)}$ {\bf models}), the
{\bf X-matter}, and the so-called {\bf Chaplygin-type gas}. A
summary of their main results is given and some difficulties
underlying the emerging dark energy paradigm are also briefly
examined.
\end{abstract}

\maketitle

\section{Introduction}

\vspace{0.1cm}
In 1998, some results based on Supernovae (SNe)
type Ia observations published independently by two different
groups, drastically changed our view about the present state of
the universe \cite{perlmutter,riess}. In brief, the Hubble-Sandage
diagram describing the observed brightness of these objects as a
function of the redshift lead to unexpected and landmark
conclusion: the expansion of the Universe is speeding up not
slowing down as believed during many decades. Implicitly, such SNe
type Ia observations suggest that the bulk of the energy density
in the Universe is repulsive and appears like a dark energy
component; an unknown form of energy with negative pressure [in
addition to the ordinary dark matter] which is probably of
primordial origin.
\begin{figure}
\vspace{.15in}
\centerline{\psfig{figure=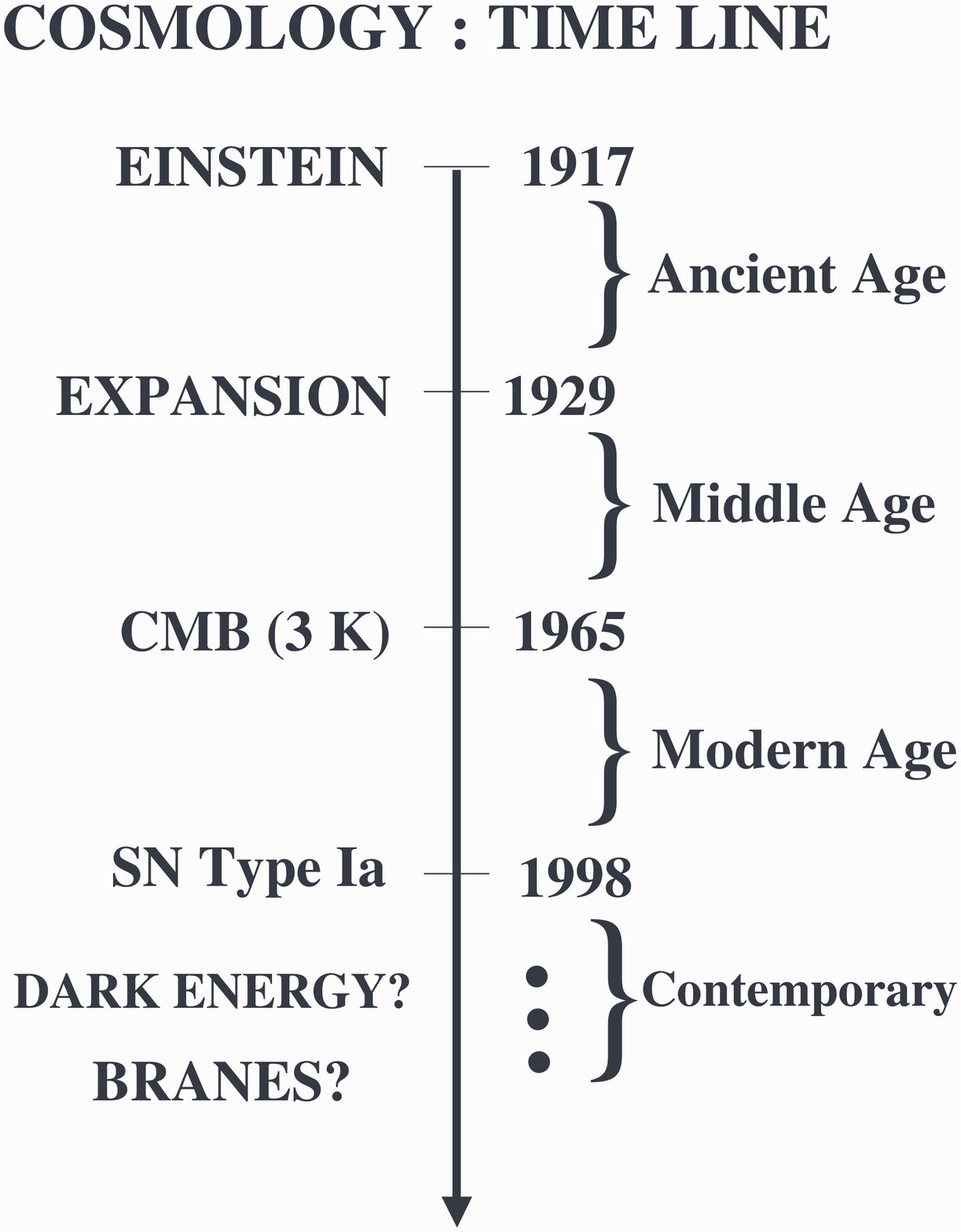,width=3.0truein,height=2.5truein}
\hskip 0.05in}
\end{figure}
In a more historical perspective, as the one shown in the
chronological scheme above, one may say that contemporary
cosmology started with the SNe ``experiments". The current
expectation is that important clues to the emerging dark energy
paradigm will be provided by the next generation of SNe projects
with advancing technology \cite{Proj}, as well as by a large set
of complementary cosmological observations.

The existence of an extra component filling the Universe has also
indirectly been suggested by independent studies based on
fluctuations of the 3K relic radiation \cite{CMB}, large scale
structure \cite{LSS}, age estimates of globular clusters or old
high redshift objects \cite{OHRG}, as well as by the X-ray data
from galaxy clusters \cite{Allen}. Actually, the angular power
spectrum of fluctuations in the cosmic microwave background (CMB)
favors a model with total density parameter $\Omega_T=1$, a value
originally predicted by inflation, whereas the density parameter
associated with cold dark matter (CDM) is $\Omega_m \sim 0.3$, a
value independently required by the power spectrum of the large
scale structure (LSS) and X-ray data from galaxy clusters (see
scheme above). The difference $\Omega_{DE} = \Omega_T - \Omega_m
\sim 0.7$ is the density parameter of the dark energy component.
Such a picture has recently been confirmed with even more
precision by the Wilkinson Microwave Anisotropy Probe \cite{WMAP},
and all these ingredients together reinforce what is usually
referred to as the standard concordance model of cosmology
\cite{OStein}.

\begin{figure}
\vspace{0.2in}
\centerline{\psfig{figure=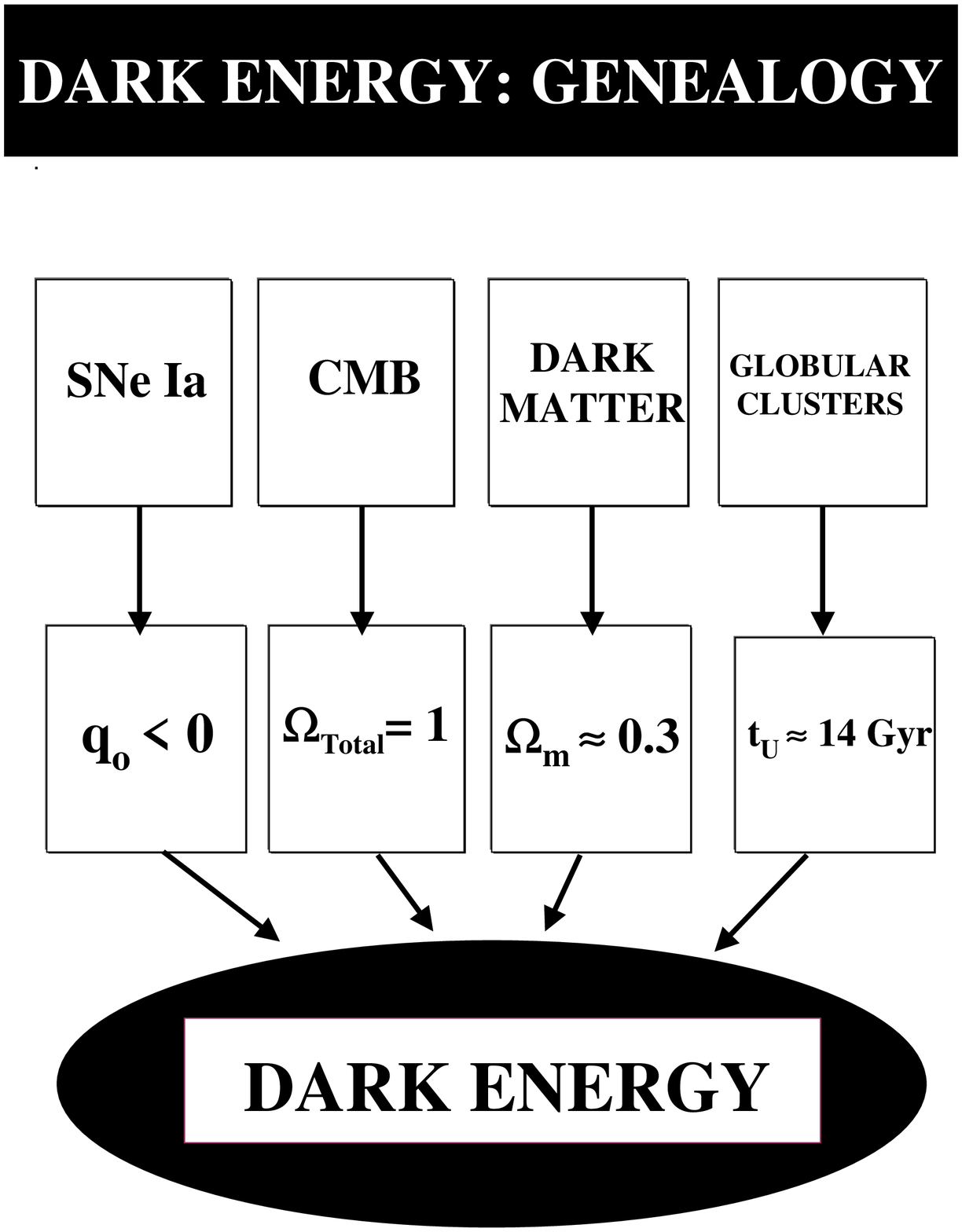,width=3.3truein,height=3.5truein}
\hskip 0.1in}
\end{figure}
Although considering that dark energy changed the traditional view
of the Universe, the absence of natural guidance from particle
physics theory about its nature gave origin to an intense debate,
as well as to many theoretical speculations. In particular, a
cosmological constant ($\Lambda$) -- the oldest and by far the
most natural candidate -- is the simplest from a mathematical
viewpoint but not the unique possibility. The $\Lambda$ term was
originally introduced by Einstein in 1917 to obtain a static world
model. It is a time independent and spatially uniform dark
component, which may classically be interpreted as a relativistic
perfect simple fluid obeying the equation of state $p_v = -
\rho_v$. In the framework of quantum field theory the presence of
$\Lambda$ is due to the zero-point energy of all particles and
fields filling the Universe which manifests itself in several
quantum phenomena like the Lamb shift and Casimir effect
\cite{Mostp}. However, there is a fundamental problem related to
such a theoretically favored candidate which is usually called the
cosmological constant problem. Shortly, it is puzzling that the
present cosmological upper bound ($\Lambda_o/8\pi G \sim 10^{-47}
GeV^{4}$) differs from natural theoretical expectations ($\sim
10^{71} GeV^{4}$) by more than 100 orders of magnitude. This
puzzle at the interface of astrophysics, cosmology, and quantum
field theory has been considered by some authors as the greatest
crisis of modern physics \cite{weinb}, and, as such, it acts like
a Damocles sword on the cosmological constant solution for the
present accelerating stage of the Universe.

Nowadays, there are many other candidates appearing in the
literature, among them:

(i) a $\Lambda(t)$-term, or a decaying vacuum energy density.

(ii) a relic scalar field (SF) slowly rolling down its potential.

(iii) ``X-matter", an extra component characterized by an equation
of state $p_{\rm x}=\omega\rho_{\rm x}$, $-1 \leq \omega < 0$.

(iv) a Chaplygin-type gas whose equation of state is given by $p=
-A/\rho^{\alpha}$, where $A$ is a positive constant and
$0\leq\alpha\leq1$.

The list is by no means as exhaustive as one may think at first
sight. Since the basic condition for an accelerating Universe is a
dominant component with negative pressure, there are other
possibilities which have occasionally been considered in the
literature \cite{mcreation}. Note also that the model dominated by
cosmological constant ($p_v = -\rho_v$) is a limiting case of the
X-matter parametrization ($\omega = -1$).

The last three candidates above (SF, X-matter, and Chaplygin gas)
share an additional physical property, namely, the effective
equation of state parameter ($\omega(z)= p/\rho$) may be a
function of the redshift. In particular, this means that many
different models may explain the same set of data. Therefore, in
order to improve our understanding of the nature of dark energy,
an important task nowadays in cosmology is to find new methods or
to revive old ones that could directly or indirectly quantify the
amount of dark energy present in the Universe, as well as
determine its effective equation of state parameter. In other
words, by learning more about the cosmic acceleration at low and
high redshifts, one may expect to discriminate among the existing
theories of dark energy by better determining $\omega$ and its
time dependence.

In this short review we present a simplified picture of the main
results and discuss briefly some difficulties of the emerging dark
energy paradigm. Since the consequences of a cosmological constant
and a rolling scalar field (usually considered the best
candidates) have already been extensively discussed in recent
review papers \cite{caldwell}, in the present work we emphasize
only the main results related to the remaining dark energy
candidates.

\section{Alternative Dark Energy Models}
\vspace{0.1cm} In what follows we restrict our attention to the
class of spacetimes described by the FRW flat line element ($c=1$)
\begin{equation}
ds^2 = dt^2 - R^{2}(t) \left[dr^{2} + r^{2}(d
 \theta^2 + \rm{sin}^{2} \theta d \phi^{2})\right]  ,
\end{equation}
where $R(t)$ is the scale factor. Such a background is favored by
the cosmic concordance model since it is a direct consequence of
the recent CMB results ($\Omega_T=1$). Now, let us discuss the
cosmic dynamics and some observational consequences for
alternative dark energy candidates.

\vspace{0.4cm}

{\large \bf A. Time-varying $\Lambda(t)$-term}

\vspace{0.4cm}

\noindent Decaying vacuum cosmologies or $\Lambda(t)$ models
\cite{bronst,ozer,freese,CW,CLW,Waga,Overduin,Lima} are described
in terms of a two-fluid mixture: a decaying vacuum medium ($\rho_v
(t) =\Lambda(t)/8\pi G,\,\, p_v = -\rho_v$) plus a fluid component
(``decaying vacuum products") which are characterized by their
energy density $\rho$ and pressure $p$. Historically, the idea of
a time varying $\Lambda(t)$-term was first advanced in the paper
of Bronstein \cite{bronst}. Different from Einstein's cosmological
constant, such a possibility somewhat missed in the literature for
many decades, and, probably, it was not important to the recent
development initiated by Ozer and Taha at the late eighties
\cite{ozer}.

The Einstein field equations (EFE) and the energy conservation law
(ECL) for $\Lambda(t)$ models are:

\begin{equation} \label{eq:rho4}
8\pi G \rho + \Lambda (t) = 3 \frac{\dot{R}^2}{R^2}\,\,,
\end{equation}
\begin{equation}
\label{eq:p4} 8\pi Gp - \Lambda (t) = -2\frac{{\ddot R}}{R}
-\frac{{\dot R}^2}{R^2}\,\, ,
\end{equation}
\begin{equation} \label{eq:ECL}
\dot \rho + 3H(\rho + p) = - {\dot \Lambda (t) \over 8 \pi G}\,\,
,
\end{equation}
where a dot means time derivative. It should be noticed that the
ECL (\ref{eq:ECL}) may be rewritten to yield an expression for the
rate of entropy production in this model as
\begin{equation}
T\frac{dS}{dt} = - \frac{{\dot\Lambda}R^{3}}{8 \pi G}\,\, ,
\end{equation}
showing that $\Lambda$ must decrease in the course of time, while
the energy is transferred from the decaying vacuum to the material
component (for more details see \cite{ozer,Lima,Lima1}).

At this point, we stress the difference between models with
cosmological constant and a decaying vacuum energy density. In the
later case, it is usually argued that the vacuum energy density is
a time-dependent quantity because of its coupling with the other
matter fields of the Universe. By virtue of the expansion, one may
suppose that the cosmological constant is relaxing to its natural
value ($\Lambda = 0$). Broadly speaking, the main goal of such
models is to determine how the energy that drove inflation at
early stages, and accelerates the universe at present is related
to the current small value of $\Lambda$. Sometimes the decaying
vacuum energy density is assumed to be an explicit time decreasing
function. However, in the majority of the papers, it depends only
implicitly on the cosmological time through the scale factor
($\Lambda \sim R^{-2}$) or the Hubble parameter ($\Lambda \sim
H^{2})$, or even a combination of them \cite{freese,CLW,Waga}. An
extensive list of phenomenological $\Lambda$-decay laws can be
seen in the paper by Overduin and Cooperstock \cite{Overduin}. All
these models have the same Achilles' heel: there is no Lagrangian
description including the coupling term (nor any physical
mechanism) governing the energy change between the decaying vacuum
and other matter fields. The expression defining $\Lambda(t)$ is
obtained either using dimensional arguments or in a completely ad
hoc way. However, although essentially phenomenological, such an
approach may indicate promising ways to solve the cosmological
constant problem by establishing the effective regime to be
provided by fundamental physics.

Certainly, one of the simplest possibilities for a decaying vacuum
energy density is $\rho_{v} = {\Lambda(t)}/{8 \pi G} = \beta
\rho_{T}$, where $\rho_{v}$ is the vacuum energy density,
$\rho_{T}=\rho_{v}+ \rho$ is the total energy density, and
$\beta\in[0,1]$ is a dimensionless parameter of order unity
\cite{freese,CLW}. By combining such a condition with the first
EFE equation one obtains the scaling law $\Lambda (t) \sim H^{2}$,
a natural result from dimensional arguments. In this scenario, the
expansion may be accelerated as required by SNe observations, and
unlike the model proposed by Ozer and Taha \cite{ozer} and Chen \&
Wu \cite{CW} for which $\Lambda \sim R^{-2}$, it solves the age
problem at $z=0$ \cite{CLW}.
\begin{figure}
\vspace{.2in}
\centerline{\psfig{figure=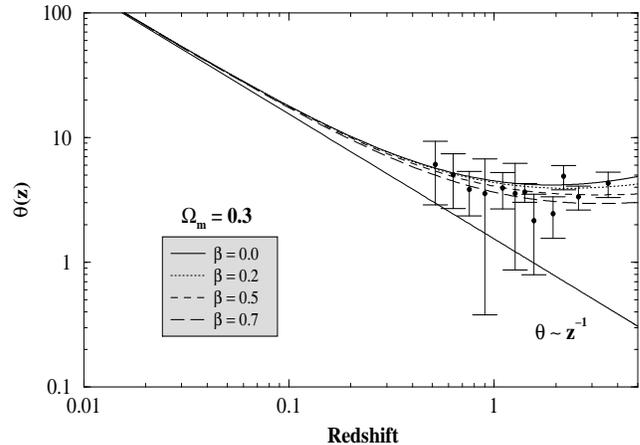,width=3.3truein,height=2.4truein}
\hskip 0.1in} \caption{The angular size - redshift relation in
decaying $\Lambda(t)$ models. The data set is composed by 145
milliarcsecond radio-sources distributed over a wide range of
redshifts ($0.011 \leq z \leq 4.72$) and binned into 12 bins
\cite{gurv}. The curves correspond to $\Omega_{\rm{m}} = 0.3$ and
a proper length $l = 26.46 h^{-1}$ pc (see \cite{CLP} for more
details).}
\end{figure}

From the observational viewpoint, $\Lambda(t)$CDM models possess
an interesting characteristic that may distinguish them from
$\Lambda$CDM models. Due to the possibility of an adiabatic photon
production the standard temperature - redshift relation may be
slightly modified. For a large class of models the temperature is
given by \cite{Lima,Lima1}
\begin{equation}
T(z) = T_o(1 + z)^{1 - \beta}\,\, ,
\end{equation}
where $T_o$ is the temperature of CMB at $z = 0$. This expression
implies that for a given redshift $z$, the temperature of the
Universe is lower than in the standard photon-conserved scenario.
Although some recent determinations of $T(z)$ (based upon the $J =
0$, 1, and 2 ground state fine-structure levels of CI) have
obtained values roughly consistent with the standard prediction,
it is well known that such measurements must be taken as upper
limits once many other excitations mechanisms may have contributed
to the observed level populations. In particular, by considering
collisional excitations, Molaro {\it et al.} \cite{mol} found a
temperature for the CMB of $T_{\rm{CMB}} = 12.1^{+1.7}_{-3.2}$ K
at $z = 3.025$. This result implies $\beta \leq 0.22$ at
$2\sigma$. More stringent constraints are furnished by big bang
nucleosynthesis (BBN). Initially, Birkel and Sarkar \cite{BS}
obtained $\beta \leq 0.13$, whereas a slightly greater upper
bound, $\beta \leq 0.16$, was further derived by Lima et al.
\cite{LMP}. In such analyzes, it was assumed that the $\beta$
parameter has the same value during the vacuum-radiation and
vacuum-matter dominated epochs. Probably, if one relaxes this
hypothesis it will be much easier to satisfy the nucleosynthesis
constraints and solve other cosmological problems. Constraints
from SNe observations, angular diameter versus redshift relation,
gravitational lensing and other kinematic tests (assuming a
constant $\beta$ parameter) have been discussed by many authors
\cite{CLW,Waga,Bloom,W,CLP,AM}.

\vspace{0.4cm}

{\large \bf B. X-Matter}

\vspace{0.4cm}

\noindent In the Cosmological scenarios driven by X-matter plus
cold dark matter (sometimes called XCDM parametrization) both
fluid components are separately conserved \cite{turner}. The
equation of state of the dark energy component is $p_x = w (z)
\rho_x$. Unlike to what happens with scalar field motivated models
where $w(z)$ is derived from the field description \cite{ratra},
the expression of $w(z)$ for XCDM scenarios must be assumed a
priori. Usually, it varies with some power of the redshift, say,
$w(z) = w_o (1 + z)^{n}$. Models with constant $w$ are the
simplest ones and their free parameters can easily be constrained
from the main cosmological tests.

More recently, in order to detect the possibility of bias in the
parameter determination due to the imposition $\omega \geq -1$,
some authors have studied models with constant $w$ by considering
two different cases: the standard XCDM ($-1 \leq \omega < 0$) and
the \emph{extended} XCDM (also named ``phantom'' energy
\cite{leq1}) in which the $\omega$ parameter violates the null
energy condition and may assume values $ < -1$. In the case of
X-ray data from galaxy clusters, for instance, a good agreement
between theory and observations for $w
> -1$ is possible if $0.29 \le \Omega_{\rm m} \le 0.33$ ($68.3\%$
c.l.) and $\omega \le -0.55$ \cite{LCA}. These results are in line
with recent analyses from distant SNe Ia \cite{garn}, SNe + CMB
\cite{efs}, Sne + LSS \cite{perl}, gravitational lensing
statistics \cite{class} and the existence of old high redshift
objects (OHRO's) \cite{lima2}. In particular, Garnavich {\it et
al.} \cite{garn} used the SNe Ia data from the High-Z Supernova
Search Team to find $\omega < -0.55$ ($95\%$ c.l.) for flat models
whatever the value of $\Omega_{\rm{m}}$ whereas for arbitrary
geometries they obtained $\omega < -0.6$ ($95\%$ c.l.). Such
results agree with the constraints obtained from a wide variety of
different phenomena, using the ``concordance cosmic" method
\cite{wang}. In this case, the combined maximum likelihood
analysis suggests $\omega \leq -0.6$, which ruled out dark
components like topological defects (domain walls and string) for
which $\omega = -n/3$, being $n$ the dimension of the defect. More
recently, Lima and Alcaniz \cite{jsa} investigated the angular
size - redshift diagram ($\theta(z)$) models by using the Gurvits'
{\it et al.} published data set \cite{gurv}. Their analysis
suggests $-1 \leq \omega \leq -0.5$ whereas Corasaniti and
Copeland \cite{cora} found, by using SNe Ia data and measurements
of the position of the acoustic peaks in the CMB spectrum, $-1\leq
\omega \leq-0.93$ at $2\sigma$. Jain {\it et al.} \cite{jain} used
image separation distribution function ($\Delta \theta$) of lensed
quasars to obtain $-0.75 \leq \omega \leq -0.42$, for the observed
range of $\Omega_m \sim 0.2 - 0.4$ while Chae {\it et al.}
\cite{class} used gravitational lens (GL) statistics based on the
final Cosmic Lens All Sky Survey (CLASS) data to find $\omega <
-0.55^{+0.18}_{-0.11}$ (68\% c.l.). Bean and Melchiorri
\cite{bean} obtained $\omega < -0.85$ from CMB + SNe Ia + LSS
data, which provides no significant evidence for XCDM behaviour
different from that of a cosmological constant. A similar
conclusion was also obtained by Schuecker {\it et al.} \cite{schu}
from an analysis involving the REFLEX X-ray cluster and SNe Ia
data in which the condition $\omega \geq -1$ was relaxed.
\begin{table}
\caption{Limits to $\Omega_{\rm{m}}$ and $\omega$}
\begin{ruledtabular}
\begin{tabular}{lcrl}
Method& Reference& $\Omega_{\rm{m}}$& \quad $\omega$\\ \hline
\hline \\ CMB + SNe Ia:.....& \cite{turner}& $\simeq 0.3$& $\simeq
-0.6$\\
 & \cite{efs}& $\sim$& $< -0.6$\\
SNe Ia + LSS.........& \cite{perl}& $\sim$ & $< -0.6$\\
GL..........................& \cite{class}& $\sim$ & $-0.55$\\
X-ray GC................& \cite{LCA} & $\simeq 0.32$& $-1$\\ X-ray
GC $^{a}$..............& \cite{LCA} & $\simeq 0.31$& $-1.32$\\ SNe
Ia....................& \cite{garn}& $\sim$&$< -0.55$\\ SNe +
X-ray GC $^{a}$..& \cite{schu}& $\simeq 0.29$&
$-0.95^{+0.30}_{-0.35}$\\ SNe Ia + GL..........& \cite{waga1}&
$0.24$& $< -0.7$\\ OHRO's..................& \cite{lima2}&
0.3&$\leq -0.27$\\ Various...................& \cite{wang}& $0.2
-0.5$& $ < -0.6$\\ $\theta(z)$.........................&
\cite{jsa}& 0.2& $\simeq -1.0$\\ & \cite{cora}&\quad $\sim$& $<
-0.96$\\ $\Delta \theta$..........................& \cite{jain}&
0.2-0.4& $\leq -0.5$\\ CMB.......................& \cite{balbi}&
0.3& $< -0.5$\\ CMB + SNe + LSS.& \cite{bean}& 0.3& $< -0.85$\\
CMB + SNe + LSS.& \cite{hann}& $\sim$& $< -0.71$\\ CMB + SNe +
LSS.\footnote{extended XCDM}& \cite{hann}& $\sim$& $> -2.68$\\
\end{tabular}
\end{ruledtabular}
\end{table}

The case for extended XCDM is an interesting one. First, it was
observed that a dark component with $\omega < -1$ may provide a
better fit to SNe Ia observations than do $\Lambda$CDM scenarios
\cite{leq1}. Although having some unusual properties, this
``phantom" behavior is predicted by different approaches as, for
example, kinetically driven models \cite{chiba1} and some versions
of brane world cosmologies \cite{sahni} (see also \cite{mcI} and
references therein). In actual fact, the best-fit model is
considerably modified when the ``phantom" behavior is allowed. In
particular, for the X-ray data from galaxy cluster quoted above,
it occurs for $\Omega_{\rm m} = 0.31$, $\omega = -1.32$ and
$\chi^2_{\rm min}=1.78$ \cite{LCA}. Such limits are more
restrictive than the ones obtained by Hannestad \& M\"ortsell
\cite{hann} by combining CMB + Large Scale Structure (LSS) + SNe
Ia data. At 95.4\% c.l. they found $-2.68 < \omega < -0.78$.

A summary of recent constraints on the dark energy parameter
$\omega$ is presented in Table I. As one may see there, the
estimates of $\Omega_{\rm m}$ and $\omega$ are compatible with the
results obtained from many independent methods. In general, joint
analyses involving X-ray data, gravitational lensing, OHRG's, SNe
type Ia, CMB, and other different methods are very welcome. First,
in virtue of the gain in precision as compared to studies using
only a specific set of data. The second reason, and, perhaps more
important, is that most of cosmological tests are highly
degenerate, thereby constraining only certain combinations of
cosmological parameters but not each parameter individually.

\vspace{0.4cm}

{\large \bf C. Chaplygin-type gas}

\vspace{0.4cm}

It is widely known that the main distinction between the
pressureless CDM and dark energy is that the former agglomerates
at small scales whereas the dark energy is a smooth component.
Such properties seems to be directly linked to the equation of
state of both components. Recently, the idea of a unified
description for CDM and dark energy scenarios has received much
attention. For example, Wetterich \cite{wet} suggested that dark
matter might consist of quintessence lumps while Kasuya
\cite{kasu} showed that spintessence type scenarios are generally
unstable to formation of $Q$ balls which behave as pressureless
matter. More recently, Padmanabhan and Choudhury \cite{pad}
investigated such a possibility trough a string theory motivated
tachyonic field. Another interesting attempt at unification was
suggested by Kamenshchik {\it et al.} \cite{kame} and further
developed by Bili\'c {\it et al.} \cite{bilic} and Bento {\it et
al.} \cite{bento}. It refers to an exotic fluid, the so-called
Chaplygin type gas (Cg), whose equation of state is
\begin{equation}\label{eq7}
p_{Cg} = -A/\rho^{\alpha},
\end{equation}
where $A$ and $\alpha = 1$ are positive constants. The above
equation for $\alpha \neq 1$ constitutes a generalization of the
original Chaplygin gas equation of state proposed in Ref.
\cite{bento} whereas for $\alpha = 0$, the model behaves as
$\Lambda$CDM. The idea of a Unified Dark-Matter-Energy (UDME)
scenario inspired by an equation of state like (\ref{eq7}) comes
from the fact that the Chaplygin type gas can naturally
interpolate between non-relativistic matter and negative-pressure
dark energy regimes \cite{bilic,bento}. Since in this approach
there is only one dark component beside baryons, photons and
neutrinos, some authors have termed this UDME scenario as a
Quartessence cosmology \cite{makler}.

Motivated by these possibilities, there has been growing interest
in exploring the theoretical and observational consequences of the
Chaplygin gas, not only as a possibility for unification of the
dark sector (dark matter/dark energy) but also as a new candidate
for dark energy only. The viability of such cosmological scenarios
has been confronted by many observational results and their two
free parameters have been constrained by many authors. For
example, Fabris {\it et al.} \cite{fabris} analyzed some
consequences of such scenarios using type Ia supernovae data (SNe
Ia). Their results indicate that a cosmology completely dominated
by the Chaplygin gas is favored in comparison to $\Lambda$CDM
models. Recently, Avelino {\it et al.} \cite{avelino} used a
larger sample of SNe Ia and the shape of the matter power spectrum
to show that such data restrict the model to a behaviour that
closely matches that of a $\Lambda$CDM models while Bento {\it et
al.} \cite{bento1,bert1} showed that the location of the CMB peaks
imposes tight constraints on the free parameters of the model.
More recently, Dev, Alcaniz \& Jain \cite{dev} and Alcaniz, Jain
\& Dev \cite{jailson} investigated the constraints on the C-gas
equation of state from strong lensing statistics and high-$z$ age
estimates, respectively, while Silva \& Bertolami \cite{bert}
studied the use of future SNAP data together with the result of
searches for strong gravitational lenses in future large quasar
surveys to constrain C-gas models. The trajectories of statefinder
parameters \cite{stat} in this class of scenarios were studied in
Ref. \cite{stat1} while constraints involving Cosmic Microwave
Background (CMB) data, Fanaroff-Ryley type IIb radio galaxies and
X-ray data from galaxy clusters, have also been extensively
discussed by many authors either as a dark energy or in the UDME
picture \cite{bert,bert1,finelli,cunha,makler}.

\begin{figure}
\vspace{.2in}
\centerline{\psfig{figure=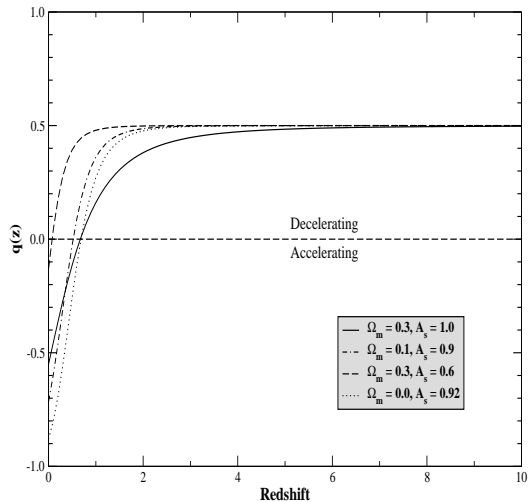,width=2.8truein,height=2.8truein,angle=-90}
\hskip 0.1in} \caption{The Chaplygin gas solution for SNe
observations. The plot shows the deceleration parameter in the
original Chaplygin gas ($\alpha=1$) as a function of redshift for
some selected values of $\Omega_{\rm{m}}$ and $A_s =
A\rho_{o}^{-2}$. The horizontal line ($q_o = 0$) divides models
with a decelerating or accelerating expansion at a given redshift.
Note that all models are accelerating at redshifts $z \lesssim 1$
(from Dev, Jain and Alcaniz \cite{dev}).}
\end{figure}

\section{Conclusion}
\vspace{0.1cm} The search for cosmologies driven by dark energy is
presently in vogue. The leitmotiv is the observational support for
an accelerated Universe provided by the type Ia supernovae (SNe)
experiments at intermediate and high redshifts.

This short review focused on some alternative candidates to dark
energy. This ubiquitous component plus the dark matter are
responsible for nearly 95\% of the matter-energy content filling
the Universe. However, different from dark matter, the extra dark
(energy) component is intrinsically relativistic and its negative
pressure is required by the present accelerating stage of the
Universe. Its tiny density and weak interaction presumably
preclude the possibility of identification in the terrestrial
laboratory. Unfortunately, even considering that we are in the
golden age of empirical cosmology, the existing data are still
unable to discriminate among the different dark energy candidates,
thereby signaling that we need better observations in order to
test the basic predictions. This means that the determination of
cosmological parameters will continue to be a central goal in the
near future. The fundamental aim is to shed some light on the
nature of the dark energy, but it is not clear if it can be
revealed using background tests with basis only in a different
equation of state. Another possibility is to add some hypothesis
concerning the nature itself (is it formed by massive or massless
particles?), and to follow examining its consequences. It is also
worth notice that the energy of this relativistic dark component
grows in the course of an adiabatic expansion. Macroscopically,
the energy increases on account of the thermodynamic work done on
the system (negative pressure). This intriguing behavior is in
marked contrast to what happens to any component with positive or
null pressure. Naturally, a possible microscopic explanation for
such a fact is of great interest because it depends on the
intrinsic nature of the dark energy, and may also have important
consequences to the ultimate fate of the Universe.

On the other hand, since the
current models are more complicated than the Einstein-de Sitter
Universe, such a situation is somewhat uncomfortable either from
theoretical or observational viewpoints. It has also to be
admitted that none dark energy model has been successful enough to
deserve the status of a ``standard model". However, the present
time for many cosmologists is very exciting because although
preserving some aspects of the basic physical picture, the new
invisible actor (dark energy) which has not been predicted by the
standard model of Particle Physics, and is responsible for
repulsive gravity, may alter profoundly the traditional view of
space-time and matter.

\vspace{0.05cm}
\vspace{0.1cm}{\bf Acknowledgments:} This work was partially
supported by the CNPq (62.0053/01-1-PADCT III/Milenio) and  FAPESP
(00/06695-0). I am also indebted to G. Steigman, J. S. Alcaniz, N.
Pires, J. Santos, R. Silva, J. V. Cunha and R. C. Santos for many
helpful discussions.


\vspace{0.01cm}
\begin{thebibliography}{30}
\vspace{0.1cm}
\bibitem{perlmutter} S. Perlmutter {\it et al.}, Nature, {\bf{ 391}}, 51 (1998); S. Perlmutter {\it
et al.}, ApJ. {\bf{517}}, 565 (1999).
\bibitem{riess} A. Riess {\it et al.}, Astron. J. {\bf{116}}, 1009
(1998).
\bibitem{Proj} See
Web sites of the
Supernova Cosmology Project (http://supernova.LBL.gov), and High-z
Supernova Search (http://cfa-www.harvard.edu/supernova).
\bibitem{CMB} P. de Bernardis et al., Nature {\bf 404}, 955
(2000); L. Knox and L. Page, PRL {\bf 85} 1366 (2000); A. H. Jaffe
et al., PRL {\bf 86}, 3475 (2001).
\bibitem{LSS}  R. G. Carlberg {\it et al.}, ApJ. {\bf{462}}, 32 (1996); A. Dekel,  D.
Burstein and S. D. M. White, In Critical Dialogues in Cosmology,
edited by N. Turok World Scientific, Singapore (1997); P. J. E.
Peebles, in Formation of Structure in the Universe, edited by A.
Dekel and J. P. Ostriker, Cambridge UP, Cambridge (1999).
\bibitem{OHRG} J. Dunlop {\it et al.}, Nature {\bf 381}, 581 (1996); Y. Yoshii, T. Tsujimoto and K. Kawara, ApJ {\bf
507}, L133 (1998); J. S. Alcaniz and J. A. S. Lima, ApJ {\bf 521},
L87 (1999); J. S. Alcaniz, J. A. S. Lima and J. V. Cunha, MNRAS
{\bf 340}, L39 (2003).
\bibitem{Allen} G. Steigman and J. E. Felten, Space Sci. Rev. {\bf 74},
245 (1995); G. Steigman, N. Hata and J. E. Felten, ApJ {\bf 510},
564 (1999); S. W. Allen, R. W. Schmidt and A. C. Fabian, MNRAS
{\bf{334}}, L11 (2002); S. Ettori, P. Tozzi and P. Rosati, A\&A
{\bf 398}, 879 (2003).
\bibitem{WMAP} C. L. Bennett {\it et al.}, ApJ Suppl. {\bf 148}, 1 (2003). See the papers of the WMAP Collaboration:
astro-ph/0302207-09, 13-15, 22-25.
\bibitem{OStein} J. P. Ostriker and P. Steinhardt, Science {\bf
300}, 1909 (2003).
\bibitem{Mostp} M. Bordag, U. Mohideen and V. M. Mostepanenko,
Phys. Rep. {\bf 353}, 1 (2001).
\bibitem{weinb} Ya. B. Zeldovich, Sov. Phys. Usp. {\bf 11},
381 (1968); S. Weinberg, Rev. Mod. Phys. {\bf 61}, 1 (1989).
\bibitem{mcreation} J. A. S. Lima and J. S. Alcaniz, A\&A {\bf{348}}, 1 (1999); L. P. Chimento,
A. S. Jakubi and N. A. Zuccala, Phys. Rev. {\bf{D63}}, 103508
(2001); W. Zimdahl, D. J. Schwarz, A. B. Balakin and D. Pav\'on,
Phys. Rev. {\bf{D64}}, 063501 (2001); M. P. Freaza, R. S. de Souza
and I. Waga, Phys. Rev. {\bf{D66}}, 103502 (2002); K. Freese and
M. Lewis, Phys. Lett. {\bf{B540}}, 1 (2002); Z. Zhu and M.
Fujimoto, ApJ {\bf{581}}, 1 (2002).
\bibitem{caldwell} R. R. Caldwell, Braz. J. Phys. {\bf{30}}, 215 (2000);
V. Sahni and A. Starobinsky, Int. J. Mod. Phys. {\bf D9}, 373
(2000); T. Padmanabhan, Phys. Rep. {\bf 380}, 235 (2003); P. J. E.
Peebles and B. Ratra, Rev. Mod. Phys. {\bf 75}, 559 (2003).
\bibitem{bronst} M. Bronstein, Phys. Z. Sowjetunion {\bf 3} (1933).
\bibitem{ozer} M. Ozer and M. O. Taha,  Phys. Lett. {B 171}, 363 (1986);  Nucl. Phys. {\bf{B287}}
 776 (1987).
\bibitem{freese} K. Freese, F. C. Adams, J. A. Frieman and E. Mottola,  Nucl. Phys.
{\bf{B287}}, 797 (1987); M. S. Berman Phys. Rev. {\bf D43}, 1075
(1991); D. Pav\'{o}n, Phys. Rev. {\bf D43}, 375 (1991); M. O.
Calv\~ao, H.P. de Oliveira, D. Pav\'on , J. M. Salim, Phys. Rev.
{\bf D45}, 3869 (1992).
\bibitem{CW} W. Chen and Y-S. Wu, Phys. Rev. {\bf D41}, 695 (1990).
\bibitem{CLW} J. C. Carvalho, J. A. S. Lima and I. Waga, Phys. Rev.
{\bf{D46}} 2404 (1992).
\bibitem{Waga} A.-M. M.Abdel-Rahman, Phys. Rev. {\bf D45}, 3497 (1992); I. Waga, ApJ {\bf 414}, 436
(1993); J. M. F. Maia and J. A. S. Lima, Mod. Phys. Lett. {\bf
A8}, 591 (1993); A. Beesham, Phys. Rev. {\bf D48}, 3539 (1993); J.
A. S. Lima and J. M. F. Maia, Phys. Rev {\bf D49}, 5597 (1994); A.
I. Arbab and A. M. M. Abdel-Rahman, Phys. Rev. {\bf D50}, 7725
(1994); J. Matygasek, Phys. Rev. {\bf D51}, 4154 (1994).
\bibitem{Lima} J. A. S. Lima and M. Trodden, Phys. Rev. {\bf D53},
4280 (1996); J. A. S. Lima, Phys. Rev. {\bf D54}, 2571 (1996).
\bibitem{Overduin} J. M. Overduin and F. I.
Cooperstock, Phys. Rev. {\bf{D58}}, 043506 (1998).
\bibitem{Lima1} J. A. S. Lima, A. I. Silva and S. M. Viegas, MNRAS {\bf 312}, 747
(2000).
\bibitem{mol} P. Molaro {\it et al.}, A\&A
 {\bf{381}}, L64 (2002).
\bibitem{BS} M. Birkel and S. Sarkar, Astrop. Phys. {\bf 6}, 197
(1997).
\bibitem{LMP} J. A. S. Lima, J. M. F. Maia and N. Pires,
IAU Simposium {\bf 198}, 111 (2000).
\bibitem{Bloom} L. F. Bloomfield Torres and I. Waga, MNRAS {\bf 279},
712 (1996).
\bibitem{W} R. G. Vishwakarma, CQG {\bf 17}, 3833 (2000); CQG {\bf 18}, 1159 (2001).
\bibitem{CLP} J. V. Cunha,
J. A. S. Lima and N. Pires, A\&A {\bf 390}, 809 (2002); J. V.
Cunha, J. A. S. Lima and J. S. Alcaniz, Phys. Rev. {\bf {D66}},
023520 (2002).
\bibitem{AM} J. S. Alcaniz and J. M. F. Maia, Phy. Rev. {\bf D67},
043502 (2003).
\bibitem{turner} M. S. Turner and M. White, Phys. Rev. {\bf{D56}}, R4439
(1997); T. Chiba, N. Sugiyama and T. Nakamura, Mon. Not. Roy.
Astron. Soc. {\bf{289}}, L5 (1997); J. S. Alcaniz and J. A. S.
Lima, ApJ {\bf{550}}, L133 (2001); J. Kujat, A. M. Linn, R. J.
Scherrer and D. H. Weinberg, ApJ {\bf{572}}, 1 (2002).
\bibitem{ratra} T. D. Saini, S. Raychaudhury, V. Sahni and
A. A. Starobinsky, Phys. Rev. Lett. {\bf{85}}, 1162 (2000); J. K.
Erickson et al., Phys. Rev. Lett. 88 121301 (2002).
\bibitem{leq1} R. R. Caldwell, Phys. Lett. {\bf{B545}}, 23 (2002).
\bibitem{LCA} J. A. S. Lima and J. S. Alcaniz, MNRAS {\bf 317}, 893 (2000); J. A. S. Lima, J. V. Cunha and J. S. Alcaniz, Phys.
Rev. D {\bf 68}, 023510 (2003).
\bibitem{garn} P. M. Garnavich {\it{et al.}}, ApJ {\bf{509}}, 74 (1998)
\bibitem{efs} G. Efstathiou, MNRAS {\bf{310}}, 842 (1999).
\bibitem{perl} S. Perlmutter, M. S. Turner and M. White, Phys. Rev. Lett.
{\bf{83}}, 670 (1999).
\bibitem{class} K.-H. Chae {\it et al.}, Phys. Rev. Lett. {\bf{89}}, 151301
(2002).
\bibitem{lima2} J. S. Alcaniz, J. A. S. Lima and J. V.
Cunha, MNRAS {\bf 340}, L39 (2003).
\bibitem{wang} L. M. Wang, R. R. Caldwell, J. P. Ostriker \& P. J. Steinhardt, ApJ {\bf{530}}, 17 (2000)
\bibitem{jsa} J. A. S. Lima and J. S. Alcaniz, ApJ {\bf{566}}, 15
(2002).
\bibitem{gurv} L. I. Gurvits, K. I. Kellermann and S. Frey,
A\&A 342, 378 (1999).
\bibitem{cora} P. S. Corasaniti and E. J. Copeland, Phys. Rev. {\bf{D65}},
043004 (2002).
\bibitem{jain}  D. Jain, A.
Dev, N. Panchapakesan, S. Mahajan, V. B. Bhatia, Int. J. Mod.
Phys. {\bf D12}, 953 (2003).
\bibitem{bean} R. Bean and A. Melchiorri, Phys. Rev. {\bf{D65}}, 041302 (2002)
\bibitem{schu} P. Schuecker, R. R. Caldwell, H. Bohringer, C. A. Collins
and Luigi Guzzo, A\&A {\bf 402}, 53 (2003).
\bibitem{waga1} I. Waga and A. P. M. R. Miceli, Phys. Rev {\bf{59}},
103507 (1999).
\bibitem{balbi} A. Balbi, C. Baccigalupi, S. Matarrese, F. Perrota
and N. Vittorio, ApJ {\bf{547}}, L89 (2001).
\bibitem{chiba1} T. Chiba, T. Okabe and M. Yamaguchi, Phys. Rev.  {\bf D62}, 023511 (2000)
\bibitem{sahni} V. Sahni and Y. Shtanov, JCAP, 0311, 014 (2003). See also astro-ph/0202346.
\bibitem{mcI} B. McInnes, ``On the nature of dark energy", Proceedings of the XVIIIth
Colloquium, Paris, eds. P. Brax, J. Martin and J,-P. Uzan, 265
(2002); S. M. Carroll, M. Hoffman and M. Trodden, Phys. Rev. {\bf
D68}, 023509 (2003).
\bibitem{hann} S. Hannestad and E. M\"{o}rtsell, Phys. Rev
D {\bf{66}}, 063508 (2002).
\bibitem{wet} C. Wetterich, Phys. Rev. {\bf D65}, 123512 (2002).
\bibitem{kasu} S. Kasuya, Phys. lett. {\bf B515}, 121 (2001).
\bibitem{pad} T. Padmanabhan and T. R. Choudhury, Phys. Rev. {\bf D
66}, 081301 (2002).
\bibitem{kame} A. Kamenshchik, U. Moschella and V. Pasquier, Phys. Lett. {\bf{B511}}, 265
(2001).
\bibitem{bilic} N. Bilic, G. B. Tupper and R. D. Violler, Phys.
lett. {\bf B535} 17 (2002).
\bibitem{bento} M. C. Bento, O Bertolami and
A. A. Sen, Phys. Rev. {\bf{D66}}, 043507 (2002).
\bibitem{makler} M. Makler, S. Q. de Oliveira and I. Waga, Phys.
Lett. {\bf B555}, 1 (2003).
\bibitem{fabris} J. C. Fabris, S. V. B. Gon\c{c}alves and P. E. de Souza,
``Fitting the supernova type Ia data with the Chaplygin gas",
astro-ph/0207430.
\bibitem{avelino} P. P. Avelino, L. M. G. Be\c{c}a, J. P. M. de
Carvalho, C. J. A. P. Martins and P. Pinto, Phys. Rev. {\bf{D67}},
023511 (2003).
\bibitem{bento1} M. C. Bento, O. Bertolami and A. A. Sen, Phys. Rev. {\bf{D67}}, 063003
(2003).
\bibitem{bert1} M. C. Bento, O. Bertolami and A. A. Sen, Phys.
Lett. {\bf B575}, 172 (2003); Gen. Rel. Grav. {\bf 35}, 2063
(2003).
\bibitem{dev} A. Dev, J. S. Alcaniz and D. Jain, Phys. Rev  {\bf{D67}}, 023515
(2003). See also astro-ph/0209379.
\bibitem{jailson} J. S. Alcaniz, D. Jain and A. Dev, Phys. Rev.
{\bf D67}, 043514 (2003). See also astro-ph/0210476.
\bibitem{bert} P. T. Silva and O. Bertolami, astro-ph/0303353.
\bibitem{stat}  V. Sahni, T. D. Saini, A. A. Starobinsky and U. Alam, JETP Lett. {\bf{77}},
 201 (2003).
\bibitem{stat1} V. Gorini, A. Kamenshchik and U. Moschella, Phys. Rev.  {\bf{D67}}, 063509
(2003); U. Alam, V. Sahni, T. D. Saini and A. A. Starobinsky,
MNRAS {\bf 344}, 1057 (2003).
\bibitem{finelli} D. Carturan and F. Finelli, Phys. Rev.
{\bf D68}, 103501 (2003); L. Amendola, F. Finelli, C. Burigana and
D. Carturan, JCAP 0307, 005 (2003).
\bibitem{cunha} J. V. Cunha, J. S. Alcaniz and J. A. S. Lima, Phys. Rev
{\bf D} (2004), accepted for publication. See also
astro-ph/0306319.
\bibitem{makler1} M. Makler, S. Q. de
Oliveira and I. Waga, {Phys. Rev} {\bf D68}, 123521 (2003).


\end{thebibliography}

\end{document}